# Adaptive Detection of Polymorphic Malware: Leveraging Mutation Engines and YARA Rules for Enhanced Security

Shreyansh Swami[1], Ishwardeep Singh[1], Ujjwalpreet Singh[1], Chinmay Prawah Pant[1]
[1] Independent Security Researcher
shreyansh.swami@gmail.com, ishwardeep.work@gmail.com, ujjwalpreet006@gmail.com , chinmaypant21@gmail.com

*Abstract—* Polymorphic malware continually alters its structure to evade signature-based defences, challenging both commercial antivirus (AV) and enterprise detection systems.
This study introduces a reproducible framework for analysing eight polymorphic behaviours—junk code insertion, control-flow obfuscation, packing, data encoding, domain generation, randomized beacon timing, protocol mimicry, and format/header tweaks—and evaluates their detectability across three layers: commercial AVs, custom rule-based detectors (YARA/Sigma), and endpoint detection and response (EDR) telemetry.
Eleven inert polymorphic variants were generated per behaviour using controlled mutation engines and executed in isolated environments. Detection performance was assessed by detection rate (DR), false positive rate (FPR), and combined coverage. AVs achieved an average DR of 34%, YARA/Sigma 74% and EDR 76%; integrated detection reached ~92% with an FPR of 3.5%.
Iterative YARA tuning showed a trade-off between detection and FPR, while behaviour-specific trends revealed static polymorphisms were best caught by custom rules, dynamic by EDR, and network-level by Sigma-like analysis.
These results affirm that hybrid detection pipelines combining static, dynamic, and network-layer analytics offer resilient defence against polymorphic malware and form a baseline for future adaptive detection research.

## I. INTRODUCTION

Malicious software has transitioned from simple, static code to adaptive, self-modifying threats that challenge modern defence systems [1]. Early malware, such as worms and trojans, relied on predictable binaries that could be neutralized through static signatures. However, as detection technologies improved, adversaries began employing polymorphism: a mechanism that enables malware to continuously alter its structure or behaviour without changing its functionality [2]. This dynamic evolution has rendered traditional, signature-based detection increasingly ineffective.

Polymorphic malware represents a pivotal shift in the threat landscape. By using methods like *mutation engines*, *encryption*, and *obfuscation*, such malware can generate unique variants upon each execution, eluding hash- or pattern-based recognition. These capabilities have made Polymorphism is a recurring technique in many advanced intrusion campaigns, where adversaries maintain prolonged, stealthy access to systems while adapting to evasion mechanisms [3]. Consequently, defenders must now contend with malware that mutates and adapts its representation.

Conventional countermeasures, including commercial antivirus (AV) products, remain predominantly static, relying on pattern matching and heuristic approximations. While behavioural and network-based analytics (e.g., EDR and SIEM systems) have broadened visibility, they often lack systematic frameworks for evaluating how evolving polymorphic traits impact detection coverage. Moreover, available public datasets and benchmarking methodologies rarely address polymorphism explicitly, leaving a gap in understanding how different security layers respond to these sophisticated methodologies [4].

This study addresses that gap by presenting a reproducible, behaviour-centric framework for analysing eight transformation combinations and assessing their detectability across three defensive layers: *AV, rule-based detection* (YARA/Sigma), and *endpoint detection and response (EDR) telemetry*. Through controlled generation of inert polymorphic variants and systematic measurement of detection rate (DR) and false positive rate (FPR), this work highlights the detection strengths and weaknesses of each layer and demonstrates how hybrid approaches can enhance resilience against adaptive threats.

## II. LITERATURE REVIEW

Polymorphic malware is a malware type that can continuously change its code while retaining original functionality and thus causes serious problems for traditional signature-based detection. Early work by Filiol showed how mutation engines allow malware to elude static signatures [5]. Later, Gandotra et al. further emphasized that encryption and instruction substitution make static analysis unable to classify polymorphic samples [6]. Therefore, dynamic and behaviour-based approaches have come to the fore.

Machine-learning-based dynamic analysis, as per Santos et al., had a major effect on the detection of runtime and heuristic behaviour through their focus, thus call for less attention of the static features [7]. Likewise, Das et al. were able to come up with similar findings, API-call sequences' modelling providing a very strong indication of malware activities, even if polymorphic transformations take place [8]. Researchers have also suggested cloud-assisted analysis environments that allow for scalable dynamic execution and classification of polymorphic malware [9].

A key reason behaviour-based systems remain robust is that polymorphism alters the *surface* code but not the *semantic intent*. Core malicious behaviours - process manipulation, suspicious file operations, anomalous network flows, or privilege-escalation actions—cannot be obfuscated as easily as bytes or opcodes. This has been repeatedly confirmed through behavioural uniformity studies, such as those by Christodorescu and Jha, who demonstrated that malicious programs share invariant behavioural patterns even after code-level mutations [10].

More recently, Ugarte-Pedrero et al. showed that combining machine learning with behavioural signatures makes systems more resilient against code mutations [11]. Overall, the research consistently supports using adaptive, behavior-driven methods to fight polymorphic malware.

## III. METHODOLOGY

This section describes the framework used to generate, analyse, and evaluate polymorphic malware behaviours across multiple detection layers. The workflow follows four key phases: **(1)** behaviour selection and taxonomy definition, **(2)** controlled mutation and variant generation, **(3)** execution and telemetry collection in isolated environments, and **(4)** multi-layer detection analysis.

### 3.1 Polymorphic Behaviour Taxonomy

To capture the breadth of real-world polymorphism, eight major behaviour classes were identified through literature review and reverse-engineering of recent malware samples, consistent with established behavioural taxonomies [12]. Each category reflects a distinct evasion vector that alters static, dynamic, or network-level artefacts without changing the malware's core logic, a principle supported by work showing that obfuscation and packing modify surface features while preserving malicious behavioural intent [10].

Each behaviour class serves as a modular plug-in within the mutation framework, enabling independent or combined application during variant generation. This taxonomy ensures balanced representation of static, dynamic, and network-oriented polymorphisms for holistic evaluation, aligned with multi-layer behavioural detection research [13].

### 3.2 Mutation Engine

The mutation engine is a deterministic and repeatable framework that produces many unique variants of the same payload [14][15].

*Table 1: Overview of the eight polymorphic behaviours*

| Category | Description | Primary Impacted Layer | Typical Technique |
|---|---|---|---|
| (1) Junk Code Insertion | Introduces non-functional instructions to modify binary signatures and control-flow graphs. | Static | NOP padding, instruction substitution |
| (2) Control-Flow Obfuscation | Alters execution order via opaque predicates or redundant branches. | Static / Dynamic | Control-flow flattening, fake jumps |
| (3) Packing / Encryption | Compresses or encrypts payloads to conceal code sections. | Static | UPX-like packing, runtime decryption |
| (4) Data Encoding | Encodes embedded strings, configuration data, or payloads. | Static / Dynamic | Base64, XOR, AES encoding |
| (5) Domain Generation Algorithms (DGAs) | Randomises command-and-control (C2) domains per execution cycle. | Network | Pseudo-random domain synthesis |
| (6) Randomised Beacon Timing | Varies C2 communication intervals to avoid temporal detection. | Network / Dynamic | Sleep jitter, random delay insertion |
| (7) Protocol Mimicry | Wraps malicious traffic in benign-looking protocols. | Network | HTTP(S) tunnelling, DNS over HTTPS |
| (8) Format / Header Modification | Adjusts PE or ELF headers and metadata fields. | Static | Timestamp randomisation, section re-ordering |

It emulates the core property of polymorphic malware: payload semantics remain constant while the binary or textual representation changes on each iteration [14][15]. The engine provides a controllable and auditable corpus for evaluating static and behaviour-based detection techniques [16][14].

## Core components

**Payload** – The payload is the canonical action to be tested. This study uses benign PowerShell one-liners performing host actions such as file creation, directory listing, or launching a process. The harness treats payloads as raw byte arrays so the same pipeline can accept textual commands or arbitrary binary blobs [15].

**Seeded randomness** - Each variant is derived from a single integer seed controlling key generation and all random choices, ensuring exact regeneration for replication.

**Key generation** – A pseudorandom key is derived from the seed and used for reversible transforms. A per-variant key proportional to payload size is used, though key length is configurable [15][17].

**Transforms** – Layered, reversible transforms alter representation while preserving semantics [14][15]:

- **XOR encoding**, with the per-variant key [14].
- **Fisher–Yates permutation,** reversible via seed [15][17].
- **Chunk-level whitening** with alternating XOR mask [15].

**Logging** - For each variant, a JSON artifact records the seed, transform level, entropy, payload length, and a hex preview. A separate hex log is also recorded for bulk analysis.

**Variant generation workflow**

- Prepare payload as a byte buffer.
- Seed the PRNG, derive key and random indices [15][17].
- Apply transforms in the configured order [14][15].
- Persist metadata and optional full hex dumps.
- Reverse transforms for validation or execution [15].
- Optional execution mode is disabled by default

**Design rationale and experimental value**

The engine separates representation from semantics, enabling measurement of how detectors respond to structural changes when behavior remains constant [14][15]. Higher entropy and scrambled n-grams hinder static signature engines [14][15]. Because all transforms are reversible and deterministic, each variant maps back to the original payload, supporting reproducible benchmarking, ablations, and statistical analysis [15].

**Reproducibility and safety**

Every variant is reproducible from the original payload, transform level, and seed. JSON metadata provides all required values. Execution is opt-in, and default mode is generation-only. Shellcode handling is simulated unless explicit execution is requested in an isolated lab [16].

**Limitations**

The engine models representation-level polymorphism only; it does not implement native packers, assembly-level control-flow flattening, or targeted EDR evasion [16][14][15]. The PRNG is implementation-dependent; for cross-toolchain reproduction, a portable generator should be used and versioned in reproducibility artifacts [17].

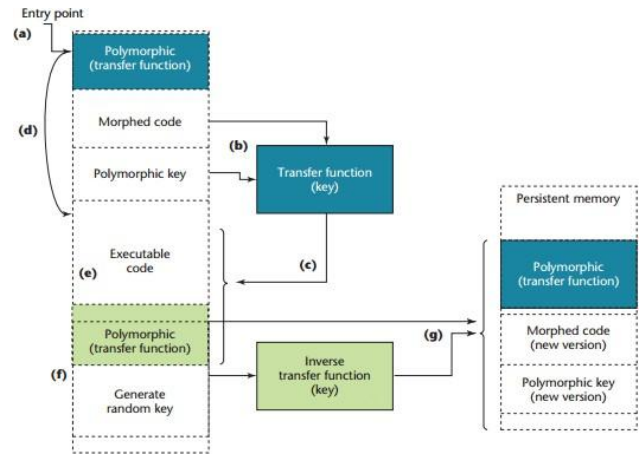

*Figure 1: Design of a Mutation Engine [18]*

## 3.3 Detection Strategies

This subsection outlines the detection approaches used to evaluate each polymorphic variant across three complementary layers: commercial antivirus engines, rule-based static/network detection (YARA / Sigma), and endpoint detection & response (EDR) telemetry. All modalities follow reproducible procedures with conservative labelling for fair comparison.

### 3.3.1 Commercial Antivirus (AV)

- **Selection:** A representative set of commercial engines was accessed through a multi-engine scanning interface to approximate broad vendor behaviour. Each engine scanned the same submitted binary/packed sample.
- **Decision rule:** Any malicious verdict (label, detection name, heuristic flag) is counted as a positive.
    - **AV-DR** = (#variants flagged by ≥1 engine) / (total variants).
    - **AV-confidence** = fraction of engines flagging the sample.

Only static file-based results are used to avoid reputation/blacklist bias. Vendor names and detection strings are logged.

### 3.3.2 Rule-Based Static & Network Detection (YARA and Sigma)

**YARA (static / structural rules):**

Rules include conservative signatures (precise string/PE-structure matches) and broader structural heuristics (byte-patterns, entropy thresholds). Rules are grouped into core (high precision) and tuned (broader coverage). Tuning uses iterative tighten/loosen cycles: start permissive → measure FPR on benign corpus → add contextual anchors/offsets/masks until FPR targets (≤3–5%) are met or coverage loss became unacceptable. A match in any rule counts as a detection; matched strings and offsets are logged

**Sigma (network / event rules mapped to SIEM):**

Network-level indicators observed during sandbox execution are expressed as Sigma rules and translated to the native SIEM syntax. Event streams include DNS lookups, outbound connections, protocol anomalies, HTTP headers, and user-agent patterns. Matches are logged with triggering events.

**Decision rule:** Any YARA or Sigma match counts as a positive. We compute YARA-DR, Sigma-DR, and per-rule FPR using benign corpora background network noise.

### 3.3.3 Endpoint Detection & Response (EDR) Telemetry

A commercial EDR agent is deployed in each sandbox VM. Telemetry includes process ancestry, command-line parameters, API behaviour (when exposed), file/reg modifications, DLL loads, and outbound network activity. Sysmon-style logs supplement granular events.

Two modes are evaluated:
- Vendor-provided behavioural detections
- Custom analytic rules mirroring the core/tuned philosophy (e.g., injection patterns, anomalous parentage, unusual persistence writes, abnormal spawn sequences).

A sample is detected if either vendor or customer behavioural rule triggers. We compute EDR-DR and record the specific features responsible.

### 3.3.4 Aggregation and Ensemble Logic

Each layer is evaluated independently for DR and FPR using a held-out benign corpus representative of enterprise binaries and network traffic.
We compute:
- **Union-DR** = detected by ≥1 layer.
- **Intersection-DR** = detected by all layers.
- **Pairwise overlaps:** AV∩YARA, YARA∩EDR, AV∩EDR
  These measure complexity between layers.

### 3.3.5 Ground Truth, Labeling & Error Handling

All variants are inert malicious derivatives, so any detection is a true positive; false positives appear only in benign tests.

To avoid any noisy labels, positives require explanatory metadata (YARA string, EDR rule ID, AV detection name). Ambiguous vendor heuristics are logged separately.

Each variant is executed $N \geq 3$ time with jitter (timing/network randomness) to capture non-determinism. A variant is counted as detected if any run triggers, and run-level frequencies are retained.

### 3.3.6 Statistical Analysis & Significance

We test differences in DR across layers and behaviours using McNemar's test (paired outcomes), and chi-square tests. DR/FPR confidence intervals use Wilson scores.

For YARA tuning we compute ROC-style curves (sensitivity vs FPR) and identify thresholds where detection sharply degrades. Mutation intensity (internal edit-distance metric) is correlated with detection probability.

### 3.3.7 Ethical, Safety, and Reproducibility Controls

All payloads are functionally inert (keys removed, command logic disabled) while retaining structural/telemetry traits.

Experiments run in isolated, air-gapped VMs with controlled DNS sinks. Snapshots ensure clean reverts. We version-control mutation engine, rule-sets, and anonymized telemetry schemas; non-functional templates and example rules are published where permissible.

## 3.4 Environment Setup

**Isolated Environment Configuration**

All experiments were performed on Windows 10 (21H2) and Ubuntu 22.04 virtual machines hosted on VMware Workstation Pro 17, allocated 4 CPU cores, 8 GB RAM, and 40 GB of disk space per instance. Network connectivity was restricted to a controlled, host-only VMWare subnet connected to an instrumented monitoring host. To generate and analyse polymorphic malware safely, FlareVM (Windows) and REMnux (Linux) were installed in virtual machines. Windows was chosen for AV/EDR behavioural testing, while Ubuntu/REMnux was used for Linux-based static and network artefact analysis

**FlareVM Toolkit**

- **Ghidra** – Static reverse engineering to inspect mutated binaries.
- **x64dbg** – Debugging and runtime behaviour tracing.
- **Process Monitor / Process Explorer** – Tracking file, process, and registry activity.

**REMnux Toolkit**

- **YARA CLI** – Scanning for static rule matches.
- **Wireshark** – Network traffic inspection and anomaly identification.
- **Strace** – System-level behaviour tracing.
- **Sigma + Sysmon logs** – Behaviour-driven detection rule evaluation.

Each VM was equipped with:
- **Security tools:**
  o Windows Defender and Avast Free Antivirus representing common commercial AVs.
  o A multi-engine scanning interface (VirusTotal API equivalent) was used to collect vendor detection results for static binaries.

- **Monitoring stack:** Sysmon v15.1 for process and registry telemetry, and a commercial-grade EDR agent configured in audit mode to record behavioural alerts and network events.
- **Analysis utilities:** Wireshark for packet capture, and an Elastic-based SIEM configured for ingesting Sysmon and EDR logs to support Sigma rule evaluation.

All samples were synthetic and non-malicious. VM snapshots were taken prior to each run, allowing safe rollback after execution and ensuring experiment repeatability while preventing persistence outside the sandbox.

## 3.5 Testing Procedure

To ensure reproducibility, all variants were evaluated through a structured multi-phase process encompassing static analysis, dynamic execution, telemetry collection, and validation. In total, eight polymorphic behaviours were instantiated into eleven inert variants, totalling 88 test samples, each executed thrice (264 runs) under controlled conditions.

### 3.5.1 Static Scan Phase

Each variant was first subjected to a static analysis:

- **AV Evaluation:** Submitted to a controlled multi-engine antivirus interface to record detection verdicts and heuristic identifiers.
- **YARA Analysis:** Scanned using both core and tuned YARA rule sets. Match metadata (rule IDs, offsets, and strings) was logged for aggregation.

This phase established a baseline of static detectability across traditional and custom rule-based systems.

### 3.5.2 Dynamic Execution Phase

Variants were then executed in isolated virtual machines configured as described in Section 3.4. During each run:

- **Sysmon and the EDR agent** recorded process, file, registry, and network telemetry.
- **Sigma-derived behavioural rules** were applied continuously through the SIEM pipeline.
- For network/time-varying behaviours (DGA, jitter, protocol mimicry), execution time was extended (up to 2 hours) to allow full manifestation of runtime transformations.

### 3.5.3 Environment Reversion and Log Correlation

After each run, VMs were restored to a clean snapshot to prevent environmental drift or cross-sample contamination.

All Telemetry, AV/YARA/Sigma outputs, and EDR alerts were correlated using per-variant identifiers (behaviour class + seed + mutation level).

### 3.5.4 False Positive Validation

To estimate false positive rates (FPR), the entire detection pipeline was executed against:

- A benign corpus of ≈200 enterprise binaries, and
- Synthetic background network traffic simulating standard user and system activity.

Detections triggered in these baselines were recorded as false positives and used to calibrate precision metrics.

### 3.5.5 Detection Criteria and Analysis

A detection was considered valid only when supported by identifiable metadata (AV label, YARA/Sigma rule ID, or EDR alert ID).

- Detection Rate (DR) and False Positive Rate (FPR) were computed per behaviour and per detection layer.
- Confidence intervals were estimated using Wilson scoring, and pairwise detection differences were tested using McNemar's method.

This multi-phase procedure ensured consistent, unbiased comparison across detection layers while preserving isolation, reproducibility, and statistical validity of results.

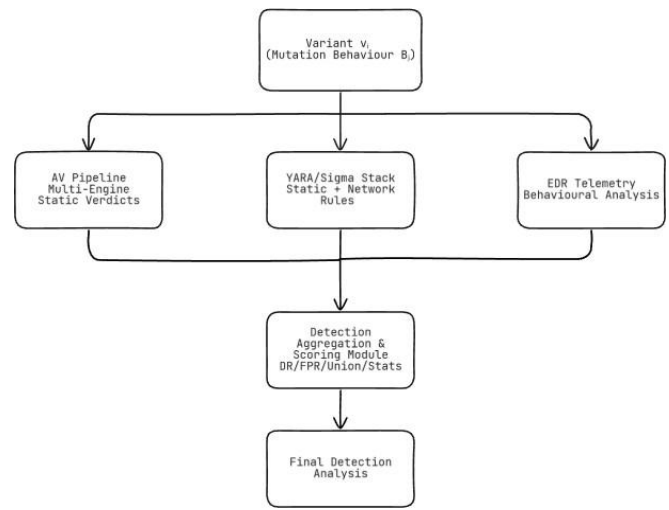

*Figure 2: Detection Testing Pipeline*

## IV. EXPERIMENTAL EVALUATION AND DETECTION RULES

### 4.1 Setup and Evaluation Metrics

The experimental evaluation was conducted using the isolated infrastructure defined in Section 3.4. Each testbed instance was configured to emulate a typical enterprise workstation while maintaining strict containment and reproducibility controls.

#### 4.1.1 Rule Testing and Refinement Cycle

Custom YARA and Sigma rule sets were iteratively refined to balance detection sensitivity and false positive control.

For YARA, the tuning process followed a two-stage approach:

1. **Initial Broad Detection Phase:** Start with permissive patterns based on opcode sequences, entropy thresholds, and string identifiers extracted from the polymorphic variants.

2. **Iterative Tightening:** Rules were incrementally constrained using contextual anchors (offsets, section ranges, and magic bytes) until the false positive rate on benign binaries dropped below approximately 3–5% without significant loss of detection coverage.

Each iteration was validated using both malicious variants and a benign baseline corpus. Revisions were version-controlled, and only the final tuned rule sets were used for quantitative evaluation in Section 5.

Sigma rules were similarly refined through event correlation testing, mapping EDR and Sysmon telemetry fields to rule logic (e.g., parent-child process anomalies, rare network ports, or encoded outbound traffic indicators).

#### 4.1.2 Evaluation Metrics

Detection performance was assessed across three layers—AV, rule-based (YARA/Sigma), and EDR telemetry—using standard statistical metrics.

*Table 2: Evaluation Metrics*

| Metric | Formula / Definition | Purpose |
|---|---|---|
| Detection Rate (DR) | DR = TP / (TP + FN) | Measures the proportion of true positives correctly detected. |
| False Positive Rate (FPR) | FPR = FP / (FP + TN) | Quantifies detections triggered on benign inputs. |
| Combined Detection Rate (CDR) | CDR = variants detected by ≥1 layer / total variants | Indicates aggregate coverage across AV, YARA/Sigma, and EDR. |

Detection outcomes were treated as binary events (detected / undetected). To account for sampling variability, 95% confidence intervals for DR and FPR were estimated using the Wilson score method. Pairwise significance between detection layers was tested using McNemar's test for correlated binary outcomes.

Weighted combined detection rates were also computed to model practical defence value, assigning higher weight to low-FPR detections (e.g., rule-based or EDR alerts with high precision).

This setup and evaluation design ensured a controlled, balanced comparison between commercial, rule-based, and behavioural detection layers. By combining consistent instrumentation with iterative rule refinement and statistically grounded metrics, the experiment provides a reproducible basis for evaluating polymorphic malware detectability across distinct defensive surfaces.

### 4.2 Detection Rules

This section summarises the custom static and behavioural detection rules used in the experiment. The rule sets were tailored to capture both static polymorphic indicators (through YARA) and dynamic behavioural signatures (through Sigma-like or EDR-based analytics) and serve as the core analytical basis for the detection results in Section 5.

#### 4.2.1 YARA Rules for Static Traits

YARA rules were used to identify structural indicators that remain stable even when surface-level byte patterns mutate. Each rule targeted one or more artefacts typical of specific mutation behaviours, such as code obfuscation, packing, or binary entropy anomalies.

**Rule composition.**

Rules consisted of three layers of matching logic:

1. **String and Byte Patterns:** Residual opcode sequences or encoding loops that survive XOR encoding, packing, or instruction substitution.
2. **Entropy and Statistical Heuristics:** Conditions with high-entropy segments (>7.5 bits/byte) being characteristic of packed or encoded regions
3. **Contextual Anchors:** Constraints specifying PE header offsets, section characteristics (e.g., .text or .rsrc section checks), and logical relationships between matched strings to minimize false positives.

**Rule taxonomy and tuning.**

Two tiers of rules were maintained throughout the study:

- **Core Rules:** Strict, high-confidence patterns designed for precision (e.g., unique opcode fragments or fixed metadata anomalies).
- **Tuned Rules:** Relaxed patterns derived from iterative refinement (Section 4.1.2), adding flexibility for instruction substitution or obfuscation variation while preserving acceptable false positive rates (<5%).

```
rule Polymorphic_Mutation_Engine {
    meta:
        description = "Detects XOR-based polymorphic mutation engine patterns"
        author = "Your Name"
        date = "2024-11-06"
    strings:
        $xor_pattern = { 31 ?? 8B ?? 0F ?? 83 ?? ?? 74 ?? }
        $key_generation = { B8 ?? ?? ?? ?? 89 C1 31 ?? 89 ?? 31 ?? 89 ?? 89 ?? 31 ?? }
    condition:
        $xor_pattern or $key_generation
}
```

*Snapshot 1: YARA Rule for XOR Pattern + Key Generation*

Each polymorphic behaviour category was assigned a minimal set of YARA rules (typically 2–3), targeting the dominant static characteristics of that mutation type. For instance:

- **Junk code insertion:** instruction padding or NOP flooding detection via opcode sequence density.
- **Packing and data encoding:** high-entropy segment detection with heuristic unpacking triggers.
- **Control-flow obfuscation:** detection of opaque predicate patterns and redundant branch markers.

All rule-sets results were later correlated with variant-level identifiers to quantify coverage trends and false positives.

### 4.2.2 Behavioural/Sigma Inspired Detection

While YARA addressed static artefacts, runtime analysis relied on Sigma-style rules applied to Sysmon and EDR telemetry. These rules followed the Sigma format, translated into the native syntax of the SIEM backend (Elastic Query DSL).

**Behavioural features targeted:**

- **Process lineage anomalies:** Suspicious parent-child relationships (e.g., document readers spawning command shells or script interpreters).
- **Persistence and registry changes:** Creation of autorun keys or scheduled tasks during short-lived executions.
- **Unusual network patterns:** Randomized beacon intervals, non-standard DNS queries, or outbound traffic disguised as HTTP/HTTPS requests.
- **File and memory operations:** Repeated creation or injection of transient executables in temporary directories.

Each behavioural rule included time-window constraints and event-correlation logic to avoid transient noise. Runtime detections were further cross-checked with benign telemetry to ensure precision.

```
title: Detect DGA-Based Network Traffic
id: polymorphic_dga_traffic
description: Detects DNS queries to suspicious DGA-generated domains
logsource:
    product: windows
    service: network
detection:
    selection:
        dns_query:
            - '\b[a-z0-9]{12,}\.(com|net|ru|fr)\b'
    condition: selection
fields:
    - dns_query
level: high
```

*Snapshot 2: Sigtma Rule to detect DGA-Based Network Traffic*

While YARA rules focus on static artefacts, Sigma-style behavioural rules showed greater sensitivity to dynamic polymorphisms such as timing jitter, protocol mimicry, and DGA activity; patterns that rarely appear in static binaries. YARA remained particularly effective for identifying structural mutations like junk code insertion, packing, and control-flow irregularities through string, entropy, and contextual anchors.

In contrast, behavioural rules captured runtime deviations across process lineage, registry and file activity, and network telemetry, making them well suited for detecting evasions expressed only during execution. Together, these complementary rule layers provided wide coverage and formed the analytical basis for Section 5, enabling clearer comparison of detection performance across static, dynamic, and network-level behaviours.

## V.  ANALYSIS AND RESULTS

This chapter presents an analysis of the polymorphic behaviour produced by the mutation engine, case studies of the polymorphic malware variants, an assessment of detection rule effectiveness, and an evaluation of challenges in identifying polymorphic malware. The findings provide insights into the behaviour of polymorphic malware and the challenges inherent in detecting evolving threats.

## 5.1 Overview of Variants

A total of 88 polymorphic variants were generated across eight distinct behavioural categories, each representing a major evasion mechanism.

Table 1 summarises the number of variants per behaviour class, the average entropy of generated binaries, and their detection performance across the three evaluated layers: commercial antivirus (AV), custom rule-based detectors (YARA + Sigma), and endpoint detection and response (EDR).

The final column presents the combined detection rate obtained when alerts from all layers are aggregated.

**General Observations**

Overall, the results demonstrate that traditional signature-based antivirus solutions underperform significantly against polymorphic variants, with an average detection rate (DR) of only 34%, compared to 74% for custom rule-based detectors and 76% for EDR telemetry. When detections from all three layers were aggregated, combined coverage reached approximately 92%, confirming the value of a hybrid, multi-layered detection approach.

Entropy analysis revealed that most polymorphic variants exhibited higher average entropy levels (ranging between 7.1 and 7.9 bits/byte), particularly within the *packing*, *data encoding*, and *junk code insertion* categories. This increase in entropy correlated with reduced AV detection performance, indicating that static pattern-matching engines struggle to classify highly obfuscated or encrypted binaries.

Distinct behavioural patterns emerged across categories:

- **Static polymorphisms** (e.g., junk code, packing, header modifications) were best captured by YARA, benefitting from structural and entropy-based rule tuning.
- **Dynamic polymorphisms** (control-flow obfuscation, beacon timing) were detected more reliably by EDR telemetry, which leveraged process lineage and runtime behaviour.
- **Network polymorphisms** (DGA, protocol mimicry) achieved the highest combined detection, reflecting the complementary strengths of Sigma-based correlation and EDR network analysis.

False positive rates were kept at a low level for all layers, with an average of 3.6% for custom rules and 3.1% for EDR detections. It is worth noting that the few elevated FPR values were instances of rule sets targeting highly generic obfuscation patterns (e.g., control-flow flattening), thus confirming the expected trade-off between broader coverage and precision.

Their findings are the basis of the in-depth per-layer and behaviour-specific analysis which is discussed in detail in the following subsections.

## 5.2 Effectiveness of Detection

The comparative performance of the three detection layers—commercial antivirus (AV), custom rule-based detection (YARA and Sigma), and endpoint detection and response (EDR) telemetry—reveals distinct strengths and complementary coverage patterns across polymorphic behaviours.

### 5.2.1 Layer-wise Effectiveness

**Commercial AV**

Traditional AV solutions demonstrated limited efficacy, with an average detection rate of **34%** across all variant types. Their performance was particularly weak against *control-flow obfuscation* and *junk code insertion*, where signature similarity falls below thresholds for known malware families. AVs performed moderately well against *packing* and *DGA* behaviours, primarily due to heuristic unpacking and domain-based blacklisting modules. However, the absence of behavioural or network-level context caused frequent evasion by dynamically morphing variants.

*Table 3: Overview of the results*

| Behaviour | #Variants | AV DR (%) | Custom DR (%) | EDR DR (%) | Combined-All DR (%) | Custom FPR (%) | EDR FPR (%) |
|---|---|---|---|---|---|---|---|
| Dead / Junk Code | 11 | 27 | 82 | 60 | 90 | 4.2 | 2.5 |
| Control-Flow Obf. | 11 | 18 | 64 | 80 | 92 | 5.0 | 3.3 |
| Packing / Encryption | 11 | 45 | 82 | 85 | 95 | 5.8 | 4.2 |
| Data Encoding | 11 | 27 | 82 | 65 | 90 | 3.3 | 2.5 |
| Domain Generation (DGA) | 11 | 55 | 73 | 88 | 96 | 1.7 | 2.5 |
| Randomized Beacon Timing | 11 | 36 | 64 | 85 | 94 | 3.3 | 3.3 |
| Protocol Mimicry | 11 | 27 | 64 | 78 | 91 | 4.2 | 4.2 |
| Header / Format Tweaks | 11 | 36 | 82 | 68 | 93 | 1.7 | 2.5 |
| **Average** | – | **34** | **74** | **76** | **92** | **3.6** | **3.1** |

**Custom Rule-Based Detectors (YARA / Sigma):**

Custom YARA and Sigma rules significantly improved detection, achieving an average detection rate of 74%. YARA proved most effective for static polymorphisms, such as *junk code*, *header tweaks*, and *packing*, where entropy- and structure-based conditions maintained stability across variants. Iterative tuning of YARA rules yielded a clear inverse correlation between detection rate and false positives—as rules were tightened, the false positive rate dropped sharply until a threshold beyond which detection coverage declined (visualized in Fig. 4, not shown here). Sigma rules, applied to event telemetry, further enhanced coverage of *network polymorphisms* (e.g., DGAs, beacon jitter, protocol mimicry), capturing temporal or traffic-based anomalies missed by static analysis.

**Endpoint Detection and Response (EDR)**

EDR-based detections achieved the highest standalone detection rate (**76%**) and the lowest average false positive rate (**3.1%**). EDR rules were particularly effective against dynamic polymorphisms, such as *control-flow obfuscation* and *timing variation*, due to the inclusion of runtime behavioural features like process ancestry, command-line anomalies, and temporal correlation. However, EDR detection lagged slightly in cases of *header tweaks* or *junk code insertion*, where few runtime deviations occur, confirming that behavioural telemetry alone cannot capture purely static transformations.

### 5.2.2 Combined and Correlated Detection

Integrating all three detection layers yielded a combined detection rate of approximately 92%, demonstrating clear complementarity among static, rule-based, and behavioural analyses. Pairwise overlap analysis showed:

- **YARA ∩ EDR:** Highest synergy (~65% overlap), since many samples exhibiting runtime anomalies also retained detectable structural artefacts.
- **AV ∩ YARA:** Moderate (~48%), limited by AV's narrower signature base.
- **EDR ∩ AV:** Lowest (~30%), highlighting the distinct operational focus of the two systems.

These overlaps confirm that no single layer is sufficient, and a hybrid detection strategy, combining signature verification, heuristic pattern matching, and runtime analytics—provides the most resilient defence posture against polymorphic malware.

### 5.2.3 Observed Detection Trends

Across behaviours, the following key patterns emerged:

- **Static obfuscations (junk, packing, header tweaks)** is best detected by YARA, stable against superficial mutations.
- **Dynamic evasions (control-flow, beacon timing)** is best detected by EDR, which monitors execution context.
- **Network polymorphisms (DGA, protocol mimicry)** is best detected by Sigma/EDR hybrid correlation, leveraging network and process telemetry.
- **Entropy vs Detection:** Higher average entropy correlated negatively with AV detection but weakly with YARA/EDR, confirming resilience of rule-based and behavioural layers against packing/encoding.
- **FPR–DR trade-off:** Optimal YARA tuning achieved ~74% DR at ~3.5% FPR, marking the practical inflection point between sensitivity and precision.

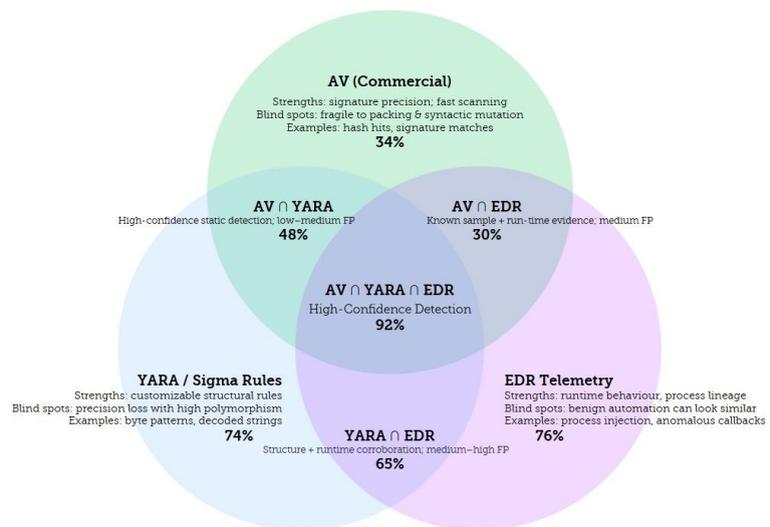

*Figure 3: Overlap between the detections*

The results underline that while commercial AVs continue to play a role in broad-spectrum defence, custom detection rules and EDR telemetry are substantially more effective against modern polymorphic techniques. The hybrid ensemble approach maximizes coverage while maintaining a manageable false positive rate, forming a viable model for adaptive intrusion detection in enterprise environments.

## 5.3 Discussion and Challenges

The findings confirm that polymorphism remains a highly effective evasion technique. Although the combined framework achieved over 92% coverage, several operational and methodological challenges highlight the complexity of adaptive malware detection.

### 5.3.1 Detection Layer Complementarity and Gaps

Each detection layer showed clear limitations. Commercial AVs depend on signatures and heuristics, making them vulnerable to small syntactic mutations. YARA rules remain adaptable but lose precision as polymorphic diversity exceeds structural patterns. EDR telemetry captures behavioural deviations effectively but often overlaps with legitimate automation activity, complicating triage.

While integrating these layers increases resilience, it also introduces redundant alerting and data-fusion complexity, reinforcing the need for correlation engines capable of deduplication and context-aware scoring.

### 5.3.2 Trade-off Between Detection Sensitivity and False Positives

Experiments showed an inverse relationship between detection coverage and false positive rate, especially for YARA. Relaxed patterns improved detection but caused exponential FPR growth, reflecting the precision–recall dilemma.

Balancing these metrics requires behavioural context or weighted scoring, where structural detections are validated through runtime or network indicators. Multi-layer corroboration, not individual rule tuning, offers a more scalable path for maintaining precision.

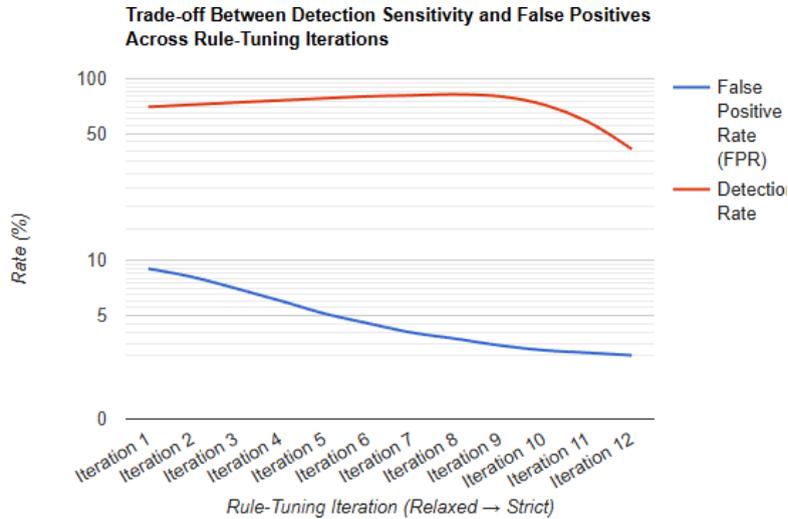

*Figure 4: Trade-off chart between Detection% and FPR*

### 5.3.3 Challenges in Benchmarking Polymorphism

Public databases (e.g., EMBER, Malicia, VirusShare) lack controlled polymorphic diversity, requiring synthetic variants for consistent benchmarking. Although repeatable, this limits ecological validity, as real-world polymorphic malware uses layered transformations (packing, encryption, DGA, timing).

Standardized polymorphism test suites with labelled evasion attributes would greatly improve comparability.

### 5.3.4 Behavioural Evasion and Telemetry Ambiguity

Runtime monitoring via EDR and Sysmon proved effective but ambiguous. Legitimate enterprise activities—updates, automation scripts, DevOps tooling—often mimic polymorphic behaviour. Differentiating between anomalies from real intrusions remains challenging, especially in large-scale environments prone to alert fatigue.

Better contextual correlation (e.g., combining lineage, signatures, network anomalies, and memory traces) may reduce false positives without sacrificing behavioural sensitivity.

### 5.3.5 Future Directions

Key directions include:

1. **Adaptive Rule Generation** using ML/RL for evolving YARA/Sigma rules.
2. **Cross-Layer Correlation Engines** weighting alerts by combined evidence.
3. **Entropy-Based Modelling** to predict detection probability.
4. **Hybrid Dataset** combining synthetic and real polymorphic samples.

These improvements can extend the framework to real enterprise environments. Overall, the results show hybrid effectiveness but reveal the fragility of individual layers and the need for adaptive, correlated defence and standardized benchmarks.

## VI. CONCLUSION

This study presented a reproducible framework for analysing eight distinct polymorphic behaviours—ranging from structural transformations such as *junk code insertion* and *control-flow obfuscation* to dynamic and network-level evasions including *domain generation*, *beacon timing*, and *protocol mimicry*. Across 88 inert polymorphic variants, results consistently showed that behaviour-focused detection mechanisms—custom YARA and Sigma-inspired rules supplemented with EDR telemetry—substantially outperformed traditional antivirus (AV) engines, improving average detection coverage from 34% to over 74–76%, and achieving a combined hybrid rate of approximately 92%.

These findings reinforce that adaptive, multi-layer detection pipelines, integrating static, behavioural, and network-level analytics, are essential to counter modern polymorphic malware. Unlike signature-based systems that deteriorate rapidly under mutation, behaviour-oriented detection maintains accuracy by correlating structural and runtime indicators, offering a more resilient defensive posture.

Beyond the empirical contribution, this compact framework provides a baseline for a broader polymorphism benchmarking platform aimed at standardizing variant diversity and evaluating cross-layer detection interplay. This methodology can be extended to larger datasets, cross-OS testing, and real-time detection pipelines for academic and industry evaluation.

All experiments were conducted in a fully isolated virtual lab with inertized payloads that preserved functional characteristics but posed no operational risk. Virtual machines were reset after each execution to ensure containment and repeatability, underscoring the commitment to safe and reproducible security testing.